\documentclass[aip,amsmath,amssymb,reprint,]{revtex4-1}
\usepackage{graphicx}
\usepackage{amsmath}
\usepackage{dcolumn}
\usepackage{cancel}
\usepackage{bm}
\usepackage{enumitem,kantlipsum}
\usepackage[nice]{nicefrac}
\usepackage[utf8]{inputenc}
\usepackage[T1]{fontenc}
\usepackage{mathptmx}
\usepackage{dsfont}
\usepackage[dvipsnames]{xcolor}
\usepackage{soul}
\usepackage[normalem]{ulem}
\usepackage{etoolbox}
\usepackage{tikz}
\usepackage{wasysym}

\begin{document}
\preprint{AIP/123-QED}
\title[Polymer single-file Dynamics]{Polymer (imperfect) single-file diffusion: A phase diagram}
\author{H. Y. Wang}
\author{Gary W. Slater}
\affiliation{Department of Physics, University of Ottawa,Ottawa, Ontario K1N 6N5, Canada}
\date{\today}
\begin{abstract}
We use Langevin dynamics (LD) simulations to investigate single-file diffusion (SFD) in a dilute solution of flexible linear polymers inside a narrow tube with periodic boundary conditions (a torus). The transition from SFD, where the time ($t$) dependence of the mean-square displacement scales like $\langle x^2 \rangle \sim t^{\nicefrac{1}{2}}$, to normal diffusion with $\langle x^2 \rangle \sim t$, is studied as a function of the system parameters, such as the size and concentration of the polymer chains and the width of the tube.  We propose a phase diagram describing different diffusion regimes. In particular, we highlight the fact that there are two different pathways to normal long-time diffusion. We also map this problem onto a one-dimensional Lattice Monte Carlo model where the diffusing object represents the polymer's center of mass. Possible extensions of this work to polydisperse polymer solutions, one-dimensional electrophoresis and DNA mapping are discussed. 
\end{abstract}
\maketitle

\section{\label{sec:intro}Introduction}

Single-file diffusion (SFD) generally refers to the Brownian motion of hard-core particles in a narrow channel, whose cross-section is too small to allow two adjacent particles to switch positions. Since the initial order of the particles cannot change with time, diffusion is constrained and the asymptotic (long-time) mean-square displacement $\langle x^2(t) \rangle$ of a given particle does not scale linearly with time $t$. The original SFD concept was proposed decades ago to describe phenomena such as ion transport through pores in biomembranes\cite{SFD_first_intro}. More recently, the SFD concept has been used to understand molecular diffusion inside zeolites\cite{SFD_Application_zeolites}, biological systems such as blood vessels\cite{ref:bloodvessel1,ref:bloodvessel2}, and molecular dynamics inside nanotubes\cite{ref:nanotubes1}.

In a pure SFD regime, the long-time mean-square displacement scales $\langle x^2(t) \rangle \sim t^{\nicefrac{1}{2}}$. Two different effects can limit the duration of the SFD regime. First, if particles can switch position (e.g., if the tube width is slightly more than twice the particles' width), we expect a transition from $\langle x^2 \rangle \sim t^{\nicefrac{1}{2}}$ to "normal diffusion" $\langle x^2 \rangle \sim t$ at long time\cite{SFD_switch_ref_1}.  Secondly, if the particles are located inside a torus (equivalent to using periodic boundary conditions), then they will move as a group at long time, a phenomenon sometimes called compound diffusion. \cite{ref:torus,ref:Jose_paper}

In this paper, we examine the dynamics of dilute solutions of monodisperse polymer chains diffusing along the axis of a narrow tube with periodic boundary conditions. We use the simplest possible model: the Langevin Dynamics of a Rouse polymer chain. This allows us to derive some analytical results and optimize the simulation times while keeping the key physical effects. In particular, our study does not consider hydrodynamic interactions (HI). While HI are not eliminated on length scales shorter than the tube diameter, they are known to be screened over longer distances\cite{ref:Dorfman_tube_hydro,ref:Dolye_intermole}.

Unlike hard particles, polymer chains can switch position even if the tube is narrow; however, this process means that two chains must become entangled in order to share the same tube Section, then possibly diffuse together before disentangling and moving away from each other. Thus, we expect the duration of the SFD regime to strongly depend on the properties of the polymer molecules, as spontaneous entanglement of two free chains is strongly hindered by entropic factors.

\begin{figure}[!ht]
\centering
\includegraphics[width = 0.47\textwidth]{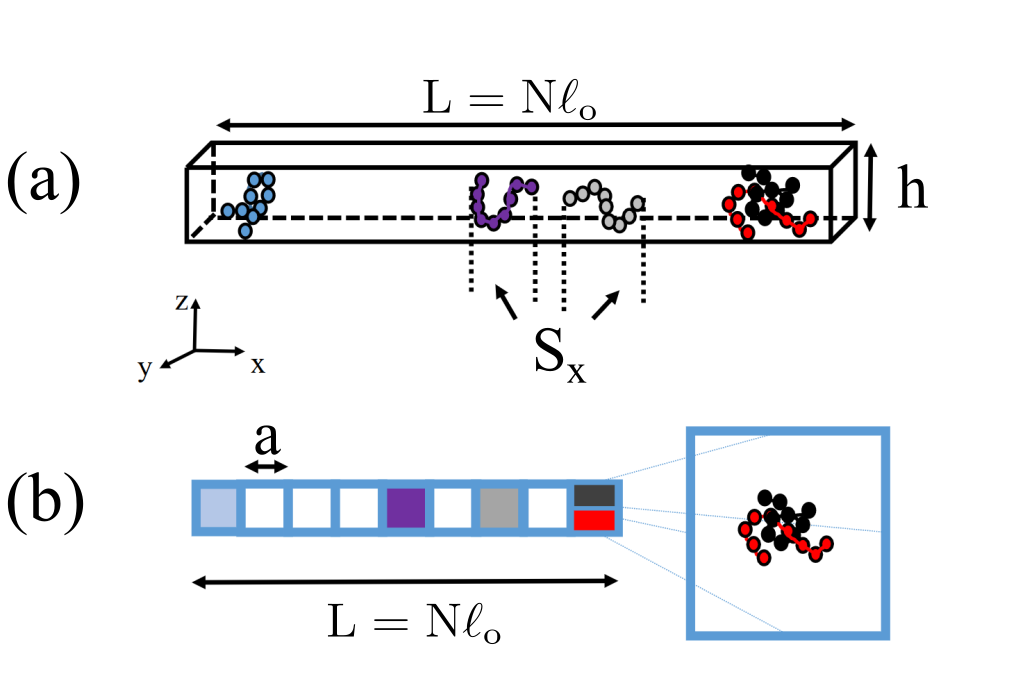}
\caption{A schematic view of the two simulation models. (a) $N$ flexible polymer chains with $M$ monomers each diffuse along a square-section tube of width $h$ and length $L$  (with periodic boundary conditions). The mean distance between the molecules is $\ell_\mathrm{o}=L/N$ while their mean axial span is $S_x$. (b) A Lattice Monte Carlo model of the same system: the cells have a size $a=S_x$ and can be either empty or occupied by one or more molecules.}
\label{fig:system_schematic}
\end{figure}

Figure \ref{fig:system_schematic}~(a) illustrates the problem at hand. A dilute solution of $N$ flexible chains (each made of $M$ monomers) is placed in a square-section tube (width $h$ and length $L=N\ell_\mathrm{o}$) with periodic boundary conditions. In principle, if the channel width $h$ exceeds twice the monomer size, two polymer chains can pass through each other. The figure also defines the mean polymer axial span $S_x(M,h)$, which is a measure of the footprint of a chain along the tube. In order to be in the dilute regime, the chains must be separated by a mean distance $\ell_\mathrm{o} > S_x$. 
Figure \ref{fig:system_schematic}~(b) shows a Lattice Monte Carlo model of the same system: the lattice cells are of size $a=S_x$ so that a polymer chain occupies a single cell. Figure~\ref{fig:system_schematic}~(b) also depicts two chains occupying the same cell, representing the entangled chains in (a). We study model (a) using Langevin Dynamics and model (b) using Monte Carlo simulations. 

This paper is organised as follows. Section \ref{sec:theory} examines both SFD theory and Rouse chain dynamics within a tube. The two simulation models are introduced in Section \ref{sec:method}. Section IV describes our results, followed by the final discussion in Section V, including possible extensions of this study.

\section{Theory}
\label{sec:theory}

\subsection{\label{subsc:hardPart}Single-file diffusion for hard particles}
\label{sec:theory_SFD}

We first consider a system of $N$ hard colloidal particles of diameter $\beta$ inside a tube of width $\beta < h < 2\beta$ and length $L=N \ell_\mathrm{o}$. Although $\ell_\mathrm{o} > \beta$ is the mean distance between the particles, the actual mean free space between two neighboring particles is $\ell=\ell_\mathrm{o} - \beta$. In fact, $\beta$ is not important here, and the system can be seen as a set of point-like objects on a line of effective length $L_e=N \ell$. We can distinguish three different regimes for this system. In the short time regime, \textit{i.e.} before the particles have diffused over a distance $\approx \ell/2$, their motion is essentially unhindered and we have
\begin{equation}
\langle x^2(t) \rangle =  2 D_\mathrm{o} t~~~~~~~~~~ t<t_\mathrm{o},
\label{eq:Short_regime}
\end{equation}
where $D_\mathrm{o}$ is the free diffusion coefficient of the particles along the tube axis. This regime persists until the particles start to collide, which occurs around time
\begin{equation}
   t_\mathrm{o} \approx (\ell/2)^2/2 D_\mathrm{o} = \ell^2/8D_\mathrm{o} ~.
\label{eq:t_o}
\end{equation}
Between the collision time $t_\mathrm{o}$ and some characteristic time $t_\lambda$, the particle is in the SFD regime with\cite{ref:Harris,ref:Levitt}
\begin{equation}
\langle x^2(t) \rangle =  2 F t^{\nicefrac{1}{2}} ~~~~~~~~~~t_\mathrm{o}<t<t_\lambda,
\label{eq:SFD_Regime}
\end{equation}
where the SFD mobility $F$ is given by
\begin{equation}
F = \ell \sqrt{\frac{D_\mathrm{o}}{\pi}} = D_\mathrm{o} \sqrt{\frac{8~t_\mathrm{o}}{\pi}}~.
\label{eq:F}
\end{equation}
If the particles are trapped in a closed circular system, the SFD regime ends when the entire population moves as a whole, like a one-dimensional Rouse chain of beads, so that
\begin{equation}
\langle x^2(t) \rangle  = 2 D_\mathrm{R}~ t ~~~~~~~~~~~~~~ t>t_\lambda,~~~
\label{eq:Rouse_regime}
\end{equation}
where 
\begin{equation}
  D_\mathrm{R} = \,D_\mathrm{o}/N  
  \label{eq:D_Rouse}
\end{equation}
is the Rouse diffusion coefficient of the population of $N$ particles. The time $t_\lambda$ and the corresponding length scale
$\lambda$ define the conditions that lead to an equality between Eqs.~\ref{eq:SFD_Regime} and \ref{eq:Rouse_regime}: 
\begin{equation}
    \lambda = 
\sqrt{\tfrac{2}{\pi}~N ~ \ell^2\,},
\label{eq:lambda}
\end{equation}
\begin{equation}
t_{\lambda} = \frac{\lambda^2}{2D_\mathrm{R}} =\tfrac{8}{\pi}\,N^2 t_\mathrm{o}.
\label{eq:t-lambda}
\end{equation}
Interestingly, $\lambda$ is essentially the geometric mean of the inter-particle spacing $\ell$ and the system effective length $L_e = N \ell$; this implies that $\lambda \ll L_e$ for large systems, so that relatively small displacements are sufficient to exit the SFD regime.

In short, we expect the diffusion sequence $D_\mathrm{o} \to F \to D_\mathrm{R}$ for such systems. However, the displacement probability distribution function remains Gaussian in all three regimes,
\begin{equation}
P(x,t) = \frac{1}{\sqrt{2 \pi \langle x^2(t) \rangle \, }} ~\exp \left( -\frac{x^2}{2\langle x^2(t) \rangle } \right),
\label{eq:Gaussian}
\end{equation}
where $\langle x^2(t) \rangle$ is given by Eq.~\ref{eq:Short_regime}, \ref{eq:SFD_Regime} or \ref{eq:Rouse_regime}.

Another regime can exist if we allow neighboring particles to switch position (for example, if the tube width $h$ is slightly larger than $2\beta$ or if we use polymer chains instead of hard particles). In essence, the effect of the collisions is then similar to that of an increased viscosity, and we expect that the mean-square displacement of a particle will be given by
\begin{equation}
\langle x^2(t) \rangle \approx \langle x^2(t_\mathrm{s}) \rangle \times \frac{t}{t_\mathrm{s}}~~~~~~~~~ t~\gg \,~t_\mathrm{s} 
\label{eq:ts}
\end{equation}
where $t_\mathrm{s}$ is the mean time between "switching" events involving a given particle and $\langle x^2(t_\mathrm{s}) \rangle$ is the distance traveled by that particle between two events. 

In the trivial case where $t_\mathrm{s} \ll t_\mathrm{o}$, Eq.~\ref{eq:Short_regime} gives $\langle x^2(t_\mathrm{s}) \rangle = 2 D_\mathrm{o}t_\mathrm{s}$ and Eq.~\ref{eq:ts} then reduces to Eq.~\ref{eq:Short_regime}: frequent switching renders their impact negligible, allowing particles to diffuse freely. 

The most interesting case is found when $t_\mathrm{o} \ll t_\mathrm{s}  \ll  t_\lambda  $. Using Eq.~\ref{eq:SFD_Regime}, we then find a new "switching" regime with
\begin{equation}
\langle x^2(t) \rangle \approx 2F t_\mathrm{s}^{\nicefrac{1}{2}} \times \frac{t}{t_\mathrm{s}} = 2 \left(\frac{F}{t_\mathrm{s}^{1/2}} \right)~t~~~~~~~~~ t_\lambda~> \,~t ~>~ t_\mathrm{s}
\end{equation}
which implies that the diffusion coefficient of a particle in the switching regime is given by
\begin{equation}
    D_\mathrm{s} = \frac{F}{t_\mathrm{s}^{\nicefrac{1}{2}}}=\left( \frac{8t_\mathrm{o}}{\pi t_\mathrm{s}}  \right)^{\nicefrac{1}{2}}~D_\mathrm{o}
\label{eq:Ds_vs_Do}
\end{equation}
The predicted diffusion coefficient is independent of the value of $N$, an expected result since collisions are local events. Using Eqs.~\ref{eq:D_Rouse}-\ref{eq:lambda}, we can rewrite this result as follows:
\begin{equation}
    D_\mathrm{s} = \left( \frac{t_\lambda}{t_\mathrm{s}}  \right)^{\nicefrac{1}{2}}~D_\mathrm{R}~.
\label{eq:Ds_vs_DR}
\end{equation}
Since $t_\lambda > t_\mathrm{s} $ in this regime, the system will first transition from free diffusion (for $t < t_\mathrm{o}$) to SFD (for $t_\mathrm{o} < t < t_\mathrm{s}$), and then from SFD to this switching regime ($t > t_\mathrm{s}$), where $D_\mathrm{s} > D_\mathrm{R}$. In fact, we expect the switching regime to extend to times $t>t_\lambda$ as well because $D_\mathrm{s} > D_\mathrm{R}$ (in other words, the Rouse regime does not exist anymore). Obviously, the sequence $D_\mathrm{o}\to F \to D_\mathrm{s}$ would also exist in an infinite system.

Finally, it is also possible to have $t_\mathrm{s} \gg t_\lambda$; this leads to
\begin{equation}
\langle x^2(t) \rangle \approx 2D_\mathrm{R}t_\mathrm{s} \times \frac{t}{t_\mathrm{s}}=2D_\mathrm{R}t~~~~~~~~~ t ~\gg~ t_\mathrm{s} .
\label{eq:ts_geq_tL}
\end{equation}
The diffusion sequence $D_\mathrm{o} \to F \to D_\mathrm{R}$ is recovered, and the particle switching process is too rare to have a major impact.

\subsection{Polymer Rouse Dynamics}

We consider a dilute solution of $N$ polymer chains (with $M$ monomers of size $b$ each) confined in a tube, as shown in Fig.~\ref{fig:system_schematic}. Free polymers can be characterized by their radius-of-gyration $\mathds{R}_g^\circ(M) \sim M^\nu$, where $\nu=3/5$ is Flory's exponent\cite{ref:Doi_Edward} . Since we use Rouse chains, the polymer diffusion coefficient is given by $D_\mathrm{o}(M)=D_\mathrm{1}/M$ \cite{ref:Doi_Edward}, where $D_\mathrm{1} \equiv D_\mathrm{o}(M=1)$ is the diffusion coefficient of one monomer. Importantly, the diffusion coefficient of a Rouse chain is still given by $D_\mathrm{o}(M)$ in a tube. Although free polymers have aspherical shapes \cite{ref:Span_asymm}, their span scales like the radius-of-gyration: $S_\mathrm{o} \sim \mathds{R}_g^\circ(M) \sim M^\nu$. 

There are two key differences between particles and polymer chains in this system: 1) The particle size $\beta$ must be replaced by the span $S_x = S_x (M,h)$; 2) polymers can in principle switch positions even if their size exceeds the tube width, $\mathds{R}_g^\circ(M) > h$. These differences will affect the various regimes described above for particles. 

\section{\label{sec:method}Methods}

In both simulation approaches, we place particles/polymer chains along a one-dimensional channel with a uniform initial separation $\ell_\mathrm{o}$. The number concentration of molecules is thus $\phi_N=1/\ell_\mathrm{o}$ and we use periodic boundary conditions. 

\subsection{Langevin Dynamics simulations}

We use a simple bead-spring model for the polymer chains. The beads are treated as point particles, and the interaction between them is the purely repulsive shifted Weeks-Chandler-Andersen (sWCA) potential\cite{ref:MCA_potential}
\begin{equation}
\frac{U_w(r)}{4\epsilon} = 
\begin{cases}
 \left( \frac{\sigma}{r}\right)^{12} - \left(
\frac{\sigma}{r} \right)^6  + \frac{1}{4}   &~r < r_\mathrm{m}\\
0 &~ r \geq r_\mathrm{m}
\end{cases}
\label{Eq:WCA}
\end{equation}
where $r$ is the distance between the two beads, $\epsilon$ is the magnitude of the potential, and $\sigma$ is the sWCA length-scale. The potential is truncated at $r=r_m \equiv 2^{\nicefrac{1}{6}} \sigma$ so that both its value and derivative are zero at this point. We use the same potential for the wall-bead interactions.

The $N$ beads of a chain are connected using the finitely extensible nonlinear elastic (FENE) potential\cite{ref:FENE_Kremer}
\begin{equation}
U_{FENE}(r) = -\tfrac{1}{2}K_{FENE}\Delta r_{max}^2~ ln\left[ 1-\left( \tfrac{r}{\Delta r_{max}}\right)^2 \right]~,
\label{Eq:Fene}
\end{equation}
with $\Delta r_{max} = 1.5 \, \sigma$ and $K_{FENE} = 30 \, \epsilon/\sigma^2$. The resulting nominal bond length, $b = 0.9679 \, \sigma$, minimizes the potential energy between two beads. Since we use a temperature $k_BT=\epsilon$, the nominal bead size $R$, which is the solution of $U_w(R)=k_BT$, is given by $R=\sigma$. 

Langevin Dynamics\cite{ref:PolymerDynamics} (LD) is used to simulate polymer dynamics. A bead is thus subjected to three types of forces: 1) a random Brownian force $\vec{F}_B$ and a dissipative force due to the fluid; 2) repulsive forces $\vec{F}_w$ near walls and other beads; 3) the FENE force that connects it to its neighbours along the chain. The corresponding equation of motion for a bead is
\begin{equation}
m \dot{\vec{v}} =  \vec{\nabla}\, U(\vec{r})+ \vec{F_B}- \zeta_\sigma \vec{v}~,
\label{Eq:Langevin}
\end{equation}
where $m$ is its mass, $\zeta_\sigma$ its friction coefficient, and $\vec{v}=\dot{\vec{r}}$ its velocity. The term $\vec{\nabla} U(\vec{r}) = \vec{\nabla} \left[ U_{w}(\vec{r})+U_{FENE}(\vec{r})\right] $ is the sum of the conservative forces (\textit{i.e.}, the FENE and sWCA forces), while the term $- \zeta_\sigma \vec{v}$ represents the damping effects of the liquid. The random force $\vec{F_B}$ is a Gaussian random variable with zero mean, $\langle \vec{F}_B \rangle =0$, and variance $\langle F_B^2 \rangle =2 \zeta_\sigma k_B T/ \Delta t$ in each spatial dimension~\cite{ref:BrowForce}. 

The LD simulations are implemented using the ESPReSo package\cite{ref:Espresso}. In the Results Section, lengths are in units of $\sigma$, $\epsilon$ acts as our unit of energy, forces are in units of $\epsilon/\sigma$, and times are in units of the Brownian time $\tau_\mathrm{o} \equiv \sigma^2/D_\sigma$ ($D_{\sigma}=k_BT/\zeta_\sigma$ is the diffusivity of a single bead). The integration time step is $\Delta t=0.01\tau_\mathrm{o}$. The polymers' diffusion coefficient along the tube axis is simply $D_\mathrm{o}(M)=D_\sigma/M=1/M$, independent of the tube width $h$. Unless otherwise indicated, ensemble averages are based on 500 to 1000 simulations each.

\subsection{Lattice Monte Carlo simulations}

\label{sec:LMC_methods}

We also simulate this system using a one-dimensional (1D) Lattice Monte Carlo method. Lattice particles replace the confined polymer chains: each of the $N$ polymer chains thus occupies a lattice cell of size $a$, which corresponds to the average axial span of a confined chain, $S_x$. The lattice has a length $L = N\ell_\mathrm{o}$ composed of $L/a$ lattice cells. One MC step (MCS) is one attempted jump per particle: this defines the unit of time $\tau_c$ ($a$ defines the unit of distance). 

At the beginning of each time step of duration $\mathrm{\tau_c}/N$, we randomly select one lattice-particle and the type of jump to be attempted: forward to the next lattice site with a probability $p_+$, backward by one site with a probability $p_-$, or stay put with a probability $p_\mathrm{o}=1-p_+-p_-$. The unbiased probabilities for a free particle are chosen to be $p_\mathrm{o} = 2/3$ and $p_\pm = 1/6$ \cite{ref:MykytaV2012}. The free diffusion coefficient is thus $D_\mathrm{o}=a^2/6~\mathrm{\tau_c}$. 

The MC algorithm should allow for polymer chains to collide, overlap/entangle, disentangle and switch position. This is equivalent to allowing two (or more) lattice particles share the same lattice site for a period of time, with probabilities that take into account the entropy cost associated with multiple occupancy. In order to reduce the number of free parameters to a strict minimum, we use only two parameters, the probabilities $P$ and $Q$, to control these polymer-polymer interactions: while $P$ controls the probability for particles to overlap, $Q$ controls the disentanglement process. 

Collision dynamics must take into account the occupancy $n_\textsf{f}$ of the site chosen as a possible final (subscript $\textsf{f}\,$) destination, as well as the occupancy $n_\textsf{i}$ of the site where the particle is initially (subscript $\textsf{i}$) located. When $n_\textsf{i}=1$ and $n_\textsf{f}=0$, the simple rules for a free particle, as described above, can be used. Our Monte Carlo algorithm thus follows these steps:

\begin{enumerate}[wide,labelindent=0pt]

\item  We start by randomly choosing one of the $N$ particles.
    
    \item Then a jump ($p_\pm$ or $p_\mathrm{o}$) is selected randomly. If $p_\mathrm{o}$ is chosen, the particle stays put.
    
    \item If a $\pm$ jump is chosen, we check the values of $n_\textsf{i}$ and $n_\textsf{f}$ to determine whether the jump will be accepted. The following tests are then applied, in this order:
    
    \begin{enumerate}

\item If $n_\textsf{i}=1$ and $n_\textsf{f}=0$, the jump is accepted.

    If $n_\textsf{i}=1$ and $n_\textsf{f} \ge 1$, the entropic cost of joining a group of $n_\textsf{f}$ molecules represents a potential barrier. We accept this jump with a probability $P^{\,n_{\textsf{f}}}$ since entropy is an extensive variable.
    
    \item If $n_\textsf{i} \ge 2$ and $n_\textsf{f} = 0$, we have three options:
    
\begin{enumerate}

    \item A probability $1/n_\textsf{i}$ that all $n_\textsf{i}$ particles move together to the chosen destination. The $1/n_\textsf{i}$ factor reflects the fact that the Rouse diffusion coefficient of a group of $n_\textsf{i}$ polymer chains is reduced by a factor $1/n_\textsf{i}$ compared to that of a single chain (i.e., $n_\textsf{i}=1$);
    
    \item  A probability $Q \le 1-1/n_\textsf{i}$ that the particle moves alone, leaving the other $n_\textsf{i}-1$ particles behind.

    \item If options i. and ii. fail, the jump is simply rejected.
    \end{enumerate}

   \item Finally, if $n_\textsf{i} \ge 2$ and $n_\textsf{f} \ge 1$, two cases are considered: 
   
   \begin{enumerate}
       \item When $n_\textsf{i} > n_\textsf{f}$ the particle does not have to overcome a potential barrier and we accept the jump with a probability $Q$ (otherwise the jump is simply rejected);
       \item If $n_\textsf{i} \le n_\textsf{f}$, on the other hand, the molecule must both leave the initial site and overcome a barrier: the jump is then accepted with a probability $Q \times P^{n_\textsf{i}-n_\textsf{f}+1}$ (otherwise the jump is simply rejected).

   \end{enumerate}

    \end{enumerate}
\end{enumerate}
We neglect the unlikely situation where $\ge 2$ particles on a lattice site move together and merge with $\ge 1$ particles on a neighboring site.

\section{\label{sec:Resul}Results}

\subsection{Langevin Dynamics: Single Polymer Chains}
As mentioned above, the axial diffusion coefficient of a confined Rouse chain with $M$ monomers should be given by $D_\mathrm{o}(M)=1/M$ in our units. Our data (not shown) give $D_\mathrm{o}(M)=0.998(2)/M$, thus confirming this prediction.

The free-solution radius-of-gyration of a polymer chain scales like $\mathds{R}_g^\circ(M) \sim M^\nu$, where $\nu$ is Flory's exponent. When confined to a narrow channel with $\mathds{R}_g^\circ(M) \gg h$, it forms a series of $n(N,h)\!=\!M/g$ non-overlapping blobs of size $\simeq h$ containing $g(h)$ monomers each\cite{ref:deGennes_2}, with $\mathds{R}_g^\circ(g) \simeq h$. The axial extension of this series of blobs is $R_{gx} \simeq n h = (\nicefrac{M}{g}) \,h$. These expressions imply that $g \sim h^{1/\nu}$, leading to $R_{gx} \sim M h^{1-1/\nu}$.

In a previous paper \cite{ref:Collision_paper}, we showed that due to finite-size effects, the Flory exponent is $\nu \approx 0.68$ in this LD model for this range of parameters, instead of $3/5$. As a result, our data gave 
\begin{equation}
\begin{split}
R_{gx}(M,h) =0.22(1)~\,(M-1)^{1.02(2)}~\,(h-2)^{-0.47(3)}
\label{Eq:size_tube}
\end{split}
\end{equation}
where we use the effective width $h-2$ instead of $h$ because the bead size is not negligible. Furthermore, the number of monomers in a blob was found to be given by 
\begin{equation}
    g(h) \approx 1.3 ~ (h-2)^{3/2}.
\label{eq:blob_start}
\end{equation}

Since the axial radius-of-gyration of a rigid rod of length $\mathcal{L}$ is $\mathcal{L}/\sqrt{12}$, a polymer chain constrained inside a tube of width $h$ can also be seen as a (porous and fluctuating) polymer rod of length $\mathcal{L}=\sqrt{12}~ R_{gx}$, when $\mathcal{L} > h$. However, we suggest that it is the axial span of the polymer chains, $S_x$, that matters here, and not their effective rod length $\mathcal{L}$, because a collision between two chains starts as soon as the first monomers touch. Therefore, we will use $S_x$ as a measure of the effective SFD axial polymer size.

\begin{figure}[!ht]
\centering
\includegraphics[width = 0.48\textwidth]{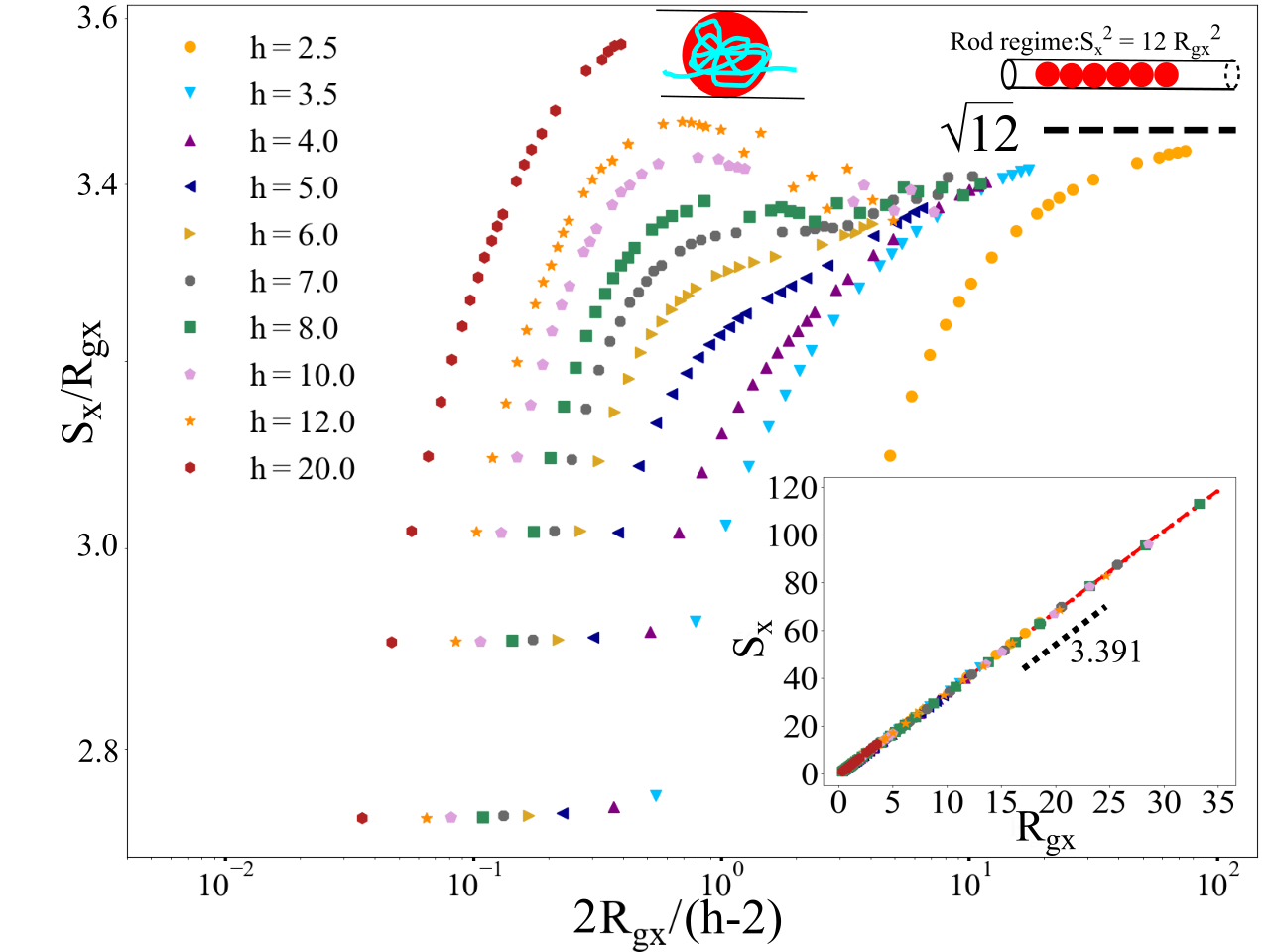}
\caption{Log-log plot of the polymer size ratio $S_x/R_{gx}$ \textit{vs} the blob size ratio $2R_{gx}/(h-2)$.  Insert: Span $S_x$ vs radius-of-gyration $R_{gx}$. The red dashed line is a linear fit with a slope of 3.391.}
\label{fig:SpanvsRg}
\end{figure}

The inset of Fig.~\ref{fig:SpanvsRg} shows that we (nearly) have a linear relationship between the axial span and the axial radius-of-gyration for all tube widths $h$ and polymer lengths $M$:
\begin{equation}
    S_x \simeq 3.391(1)\,R_{gx}~.
\label{eq:Rg_vs_Span}
\end{equation}
Equation~\ref{eq:Rg_vs_Span} is convenient, but it obscures subtle effects, as the main part of Fig.~\ref{fig:SpanvsRg} reveals. The ratio $S_x/R_{gx}$ increases towards the asymptotic value of $\sqrt{12}=3.4641$ as we increase $M$, but this increase is monotonic only for very tight tubes ($h < 7$). In fact, $S_x/R_{gx}$ reaches a maximum that exceeds $\sqrt{12}$ when $h>10$. These effects occur because a cigar-shaped polymer conformation orients under weak constraints before blob formation begins. A detailed examination of the balance between orientation and compression is outside this article's scope.

\subsection{\label{subsec:basicRes}Langevin Dynamics Results -- Polymer Chains}

We begin simulations by placing $N$ polymer chains, each with $M$ monomers, in a tube of width $h$, ensuring their centers of mass are initially separated by a distance $\ell_\mathrm{o}$ large enough to prevent contact at $t=0$. All $N$ chains were individually pre-simulated to achieve relaxed conformations. The position of the chains' CM (and other output parameters as needed) is tracked throughout the simulations, and the mean-square displacements (MSD) are computed as a function of time.

\begin{figure}[!ht]
\centering
\includegraphics[width = 0.45\textwidth]{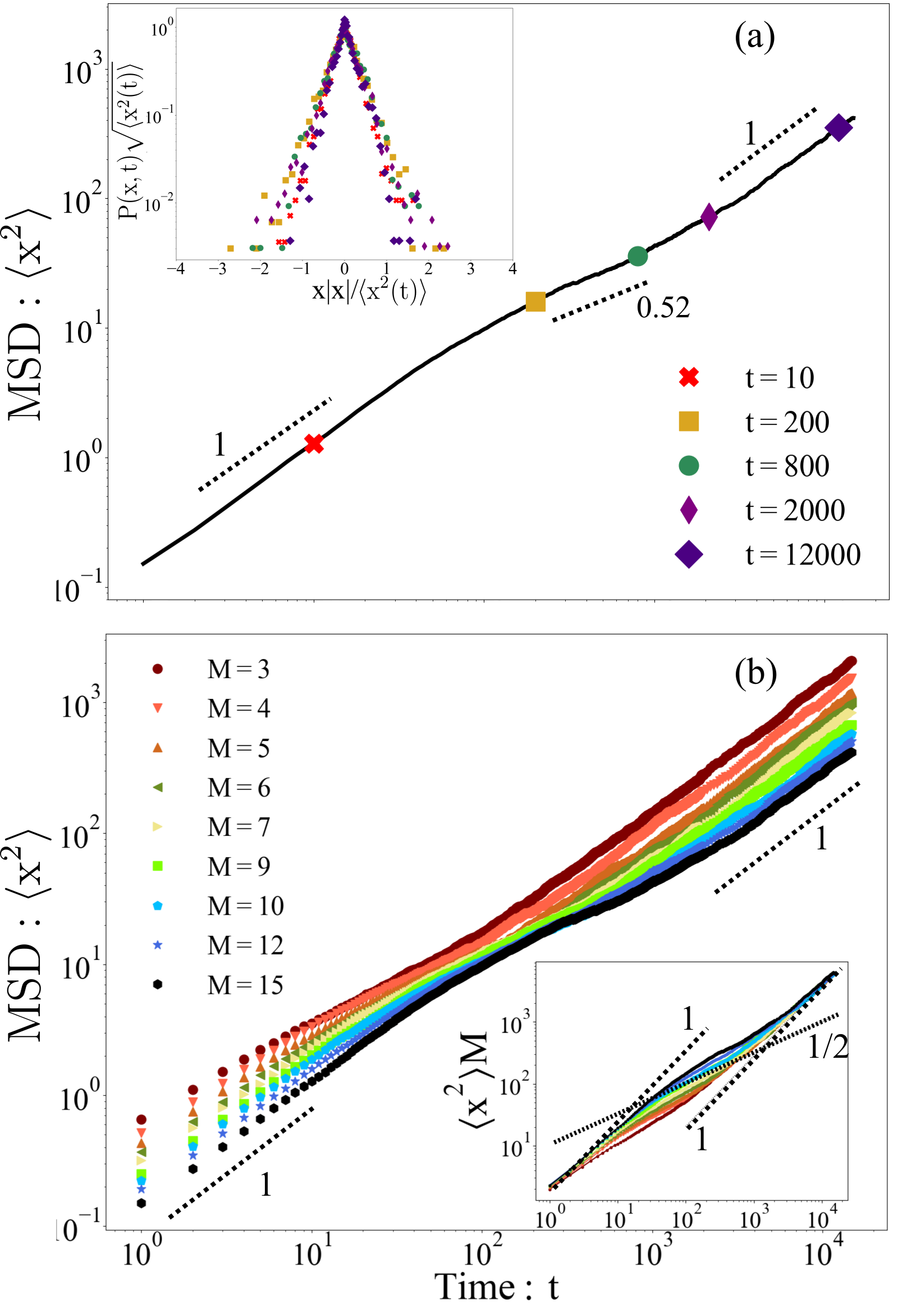}
\caption{(a) Mean-square displacement (MSD) \textit{vs} time $t$ for the $M=15$ case presented in (b). Inset: scaled probability distribution function vs scaled position for the five different times highlighted in the main plot. (b) MSD \textit{vs} time $t$ for several molecular weights $M$. The tube width is $h$ = 3.5, the number of polymer chains is $N$ = 5, and the initial separation $\ell_\mathrm{o}=M+1$. Insert: Same data with the y-axis multiplied by $M$. The two dashed lines with a unit slope differ by a factor of $N^2=25$, as predicted by theory. We also show a dotted line with a slope of $\nicefrac{1}{2}$, the value expected in the SFD regime.}
\label{fig:MSD_MD}
\end{figure}

Figure~\ref{fig:MSD_MD} (a) shows a log-log plot of the time evolution of the MSD $\left< x^2(t) \right>$ for $N=5$ polymer chains of size $M=15$ diffusing in a tube of width $h=3.5$. We clearly see the normal diffusion regimes, as predicted by Eqs.~\ref{eq:Short_regime} and \ref{eq:Rouse_regime} for short and long times respectively. We also observe the intermediate SFD regime described by Eq.~\ref{eq:SFD_Regime}. The slope in the SFD regime,  $\alpha=0.52(1)$, is consistent with the value of $\nicefrac{1}{2}$ predicted by Eq.~\ref{eq:SFD_Regime}. The inset gives the distribution function for the five different times marked in the main plot; in agreement with Eq.~\ref{eq:Gaussian}, the distribution is Gaussian in all cases.

Similar data is presented in Fig.~\ref{fig:MSD_MD} (b) for sizes $3 \le M \le 15$. Although we see the same three regimes in all cases, the intermediate regime starts earlier for smaller sizes. Because $D_\mathrm{o} \sim 1/M$ and $D_\mathrm{R} \sim N^2/M$ while $F \sim D_\mathrm{o}^{\nicefrac{1}{2}} \sim 1/M^{\nicefrac{1}{2}}$, the curves get closer to each other in the intermediate regime: the difference between the $M=3$ and $M=15$ curves is a factor of 5 at both ends but only a factor of $\sqrt{5}$ in the middle, leading to a visible narrowing. The inset shows the same data when we multiply the y-axis by $M$ in order to remove the trivial $1/M$ dependence of $D_\mathrm{o}$ and $D_\mathrm{R}$. The different data sets then overlap at short and long times, while the SFD regime forms a bubble. The gap between the two asymptotic dashed lines (with slopes of 1) is a factor of $D_\mathrm{R}/D_\mathrm{o}=N^2=25$ here.

\begin{figure}[!ht]
\includegraphics[width = 0.45\textwidth]{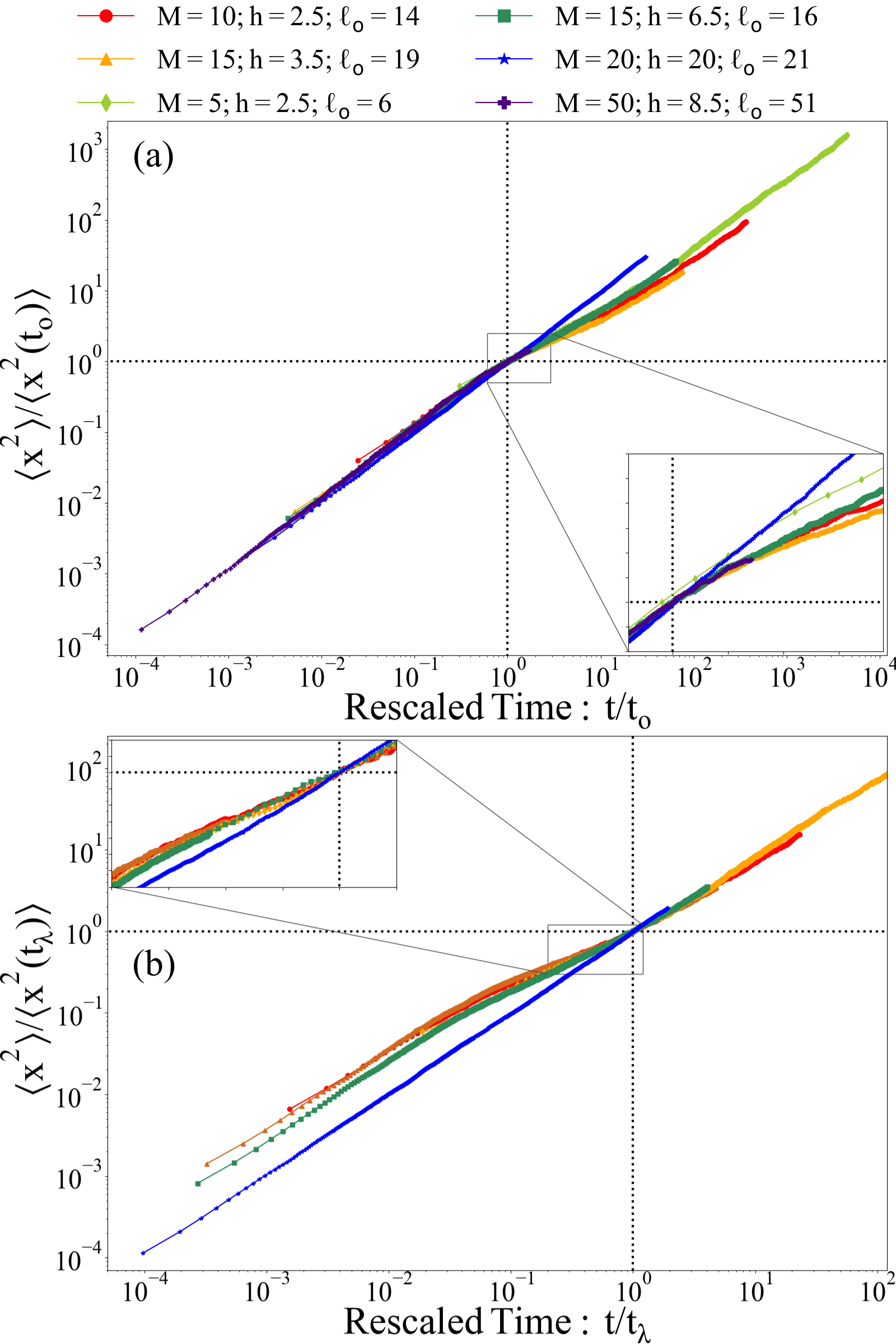}
\caption{(a) Mean-square displacement (MSD) vs time $t$ showing the transition between the free diffusion and SFD regimes. The time axis is scaled using the predicted transition time $t_\mathrm{o}$ given by Eq.~\ref{eq:SFD_touch_time} while the y-axis is scaled using the MSD values at $t=t_\mathrm{o}$. (b) Similar data showing the transition between the SFD and Rouse regimes. The time axis is scaled using the predicted transition time $t_\lambda$ given by Eq.~\ref{eq:tau_lambda_poly} while the y-axis is scaled using the MSD values at $t=t_\lambda$. } 
\label{fig:MSD_Rescale}
\end{figure}

The ratio $M/g(h)$, which gives the number of blobs formed by a chain of size $M$ in a tube of width $h$, determines how difficult it is for this chain to change position with one of its neighbors. Polymer chains rarely change position (if at all) when $M/g(h) \gg 1$, so the system then exhibits strong SFD characteristics. The time required for a chain to collide with a neighbor is given by Eq.~\ref{eq:t_o}; for the current system, it reads
\begin{equation}
    t_\mathrm{o} = \frac{\ell^2}{2D_\mathrm{o}(M)} = \frac{M(\ell_\mathrm{o}-S_x)^2}{8~D_\sigma}~.
\label{eq:SFD_touch_time}
\end{equation}
Beyond this time scale, the chains enter the SFD regime because their motion is constrained by their two neighbors and we expect $\left< x^2(t) \right> \sim t^{\nicefrac{1}{2}}$. Figure~\ref{fig:MSD_Rescale}~(a) shows that if we rescale both axes using this predicted time, the curves collapse up to $t=t_\mathrm{o}$ and diverge at longer times. The slope changes from unity to about $\nicefrac{1}{2}$ in all cases, the only exception being the $[h=20,M=20]$ system (blue stars) because the polymer chains remain in the free diffusion regime at all times. 

The SFD regime ends when the cooperative motion of the polymer chains lead to a global Rouse regime. Equation~\ref{eq:t-lambda} gives the following transition time 
\begin{equation}
  t_{\lambda} = \frac{\lambda^2}{2D_\mathrm{R}(M,N)} 
  = \frac{\tfrac{2}{\pi}~N ~ \ell^2}{2 D_\mathrm{o}(M)/N}=
\frac{M N^2 (\ell_\mathrm{o}-S_x)^2}{4 \pi \, D_\sigma}  ~.~ 
\label{eq:tau_lambda_poly}
\end{equation}
Figure~\ref{fig:MSD_Rescale}~(b) shows that if we now rescale both axes using this predicted transition time, the curves collapse at the transition point. The slope changes from about $\nicefrac{1}{2}$ to unity in all cases, with the same exception.

Polymer SFD thus shows properties similar to hard particle SFD when chain order is maintained. Next Section examines what happens when the conditions are such that the chains can and do change position, which is obviously expected to compete with SFD and cooperative Rouse dynamics.

\subsection{\label{subsec:MD_phase_switching}Polymer Chains Switching Positions}

The notion of positional exchange can be ambiguous in the context of polymers. Unlike rigid particles that possess well-defined spatial boundaries, polymer chains continuously undergo deformation and can entangle (or overlap) for extended periods of time. Our primary interest here lies in understanding how the frequency of these events varies with system parameters. In order to do this, we chose to measure the time it takes for the first such event to be observed in a simulation.

\begin{figure}[!ht]
\includegraphics[width = 0.45\textwidth]{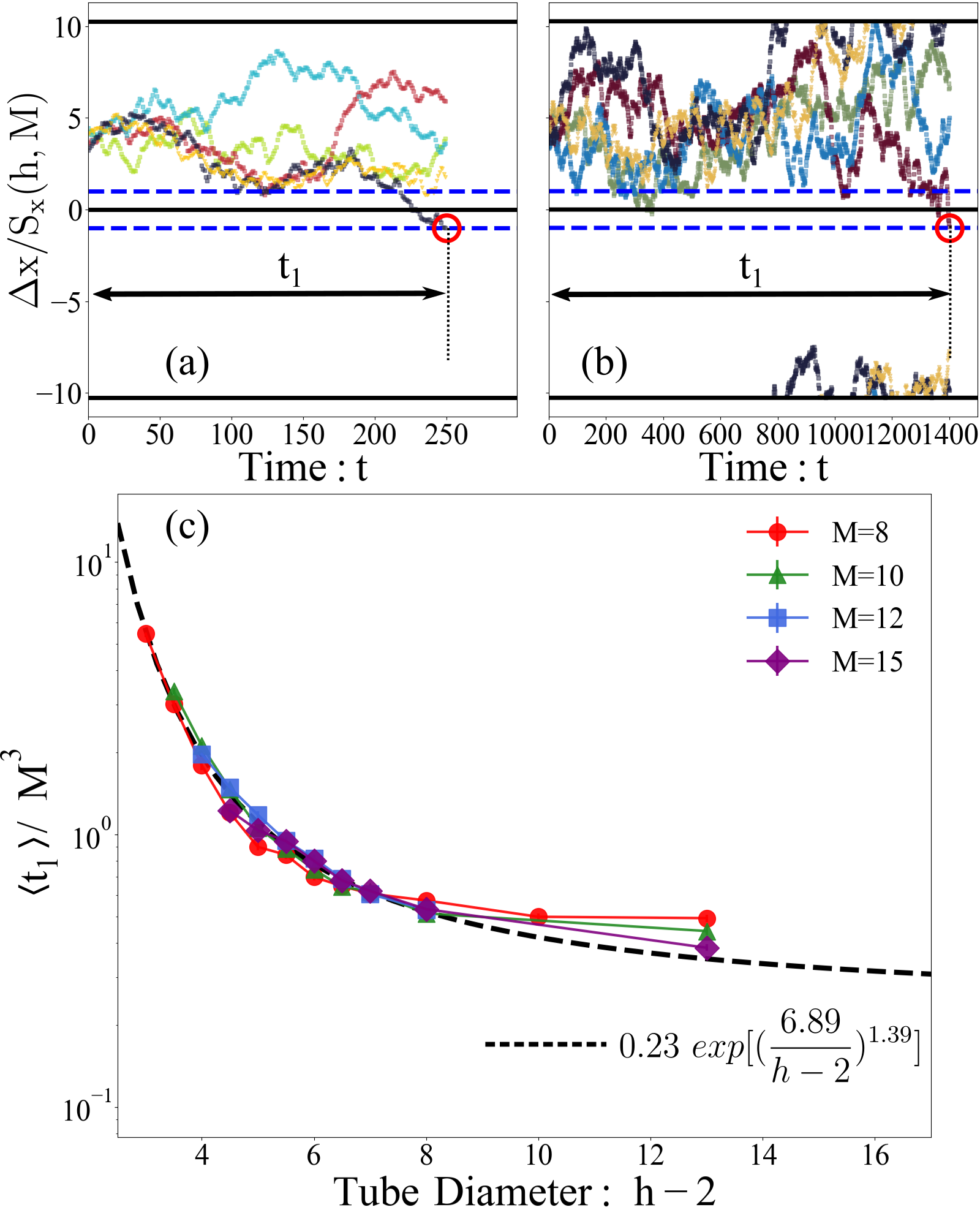}
\caption{(a)-(b) Simulation examples showing the time dependence of the scaled distance $\Delta x(t)/S_x$ between all five pairs of nearest neighbor chains. The system parameters are $N=5$, $M=8$, $h=6$, $\Delta x(0)=M+3=11$ and $S_x=2.67$. The first two polymer chains to change positions do so at times $t_1=250$ and $t_1 = 1401$. (c) Scaled mean first switching time $\langle t_1 \rangle/M^3$ as a function of the effective tube width $h-2$. The dashed line shows the fitting function Eq.~\ref{eq:SFD_swap_time}. }

\label{fig:switching_figure}
\end{figure}

We track the distance $\Delta x_{i,i+1}$ between the CMs of each pair $(i,i+1)$ of adjacent polymer chains (with $i \in 1...N-1$) as a function of time. We start the system with a uniform distance $\Delta x_{i,i+1}(0) \equiv \ell_\mathrm{o}=M+3$ between neighbors, and we record the time $t_1$ at which we observe two chains changing position for the first time. We repeat this process for 891-1000 times for each combination of tube width $h$ and chain length $M$, with $N=5$ chains in each case. In some cases, no polymer switching was observed during a simulation, hence the smaller ensemble).  Some range of parameters, for example $M=15$ and $h \le 4.5$, did not produce any chain switching event.

Figures~\ref{fig:switching_figure}~(a)-(b) show examples with $M=8$ and $h=6$; the time evolution of the $N=5$ chain-chain gaps is shown, and we see that two chains change positions at times $t_1=250$ (in A) and $t_1 = 1401$ (in B). Figure~\ref{fig:switching_figure}~(c) shows the mean time $\langle t_1\rangle$ as a function of the tube width $h$ for four different chain lengths $M$. These times diverge when $(h-2) \to 0$ because two monomers cannot be side by side in the tube. In the opposite wide tube limit, $\langle t_1 \rangle$ eventually plateaus because we simply have free diffusing chains. We can estimate the switching time in the large $h$ and $M$ limits using the simple approximation 
\begin{equation}
    \langle t_1 \rangle_{_\infty} \approx \frac{\ell_\mathrm{o}^2}{2 D_\mathrm{o}(M)}
\end{equation}
Since $\ell_\mathrm{o} =M+3$ and $D_\mathrm{o} \sim 1/M$, the mean time should then scale approximately as $\langle t_1 \rangle_{_\infty} \sim M^3$. We thus scaled the y-axis in Fig.~\ref{fig:switching_figure}~(c) accordingly, which indeed results in the near collapse of all four data sets. 

The entropy\cite{ref:deGennes_2} of a long chain  in a tube (i.e., a chain that forms blobs) scales like $S \sim M h^{-1/\nu}$. Equation \ref{eq:blob_start} indicates that we need molecules of size 
\begin{equation}
  M > g(h) =  1.3 (h-2)^{\nicefrac{3}{2}}  
\end{equation}
to be in this regime. For the four molecular weights used in Fig.~\ref{fig:switching_figure}~(c), this means tubes of width $h(M=15)<7$, $h(M=12)<6.4$, $h(M=10)<6$, and $h(M=8)<5$. Since several data points are in this regime, we fitted the four data sets for $h<7$ using an exponential function to obtain: 
\begin{equation}
    \frac{\langle t_{1} \rangle}{M^3} ~=~ 0.23(4)~ \exp \left[ \left[ {6.9(8)}/{(h-2)} \right]^{1.4(1)\,} \right]
\label{eq:SFD_swap_time}
\end{equation}
This form is compatible with the fact that the potential barrier is of entropic origin. Since we have $\nu \approx 0.68$ here due to finite size effects, this implies that $S \sim h^{-1.47}$, in fair agreement with Eq.~\ref{eq:SFD_swap_time}. The range of polymer sizes we could simulate is not large enough to fully study the M-dependence of $\langle t_1 \rangle$. The critical tube size obtained in Eq.~\ref{eq:SFD_swap_time} is $h = 8.9$, in decent agreement with our estimates above.

The theory in Section~\ref{sec:theory_SFD} for the transition regime with switching times in the range $t_\lambda~\gg \,~t_\mathrm{s} ~\gg~ t_\mathrm{o}$ predicts two properties for the asymptotic diffusion coefficient $D_\mathrm{s}$. First, Eq.~\ref{eq:Ds_vs_Do} indicates that $D_\mathrm{s} \sim t_\mathrm{s}^{-\nicefrac{1}{2}}$ because polymer switching is more efficient than Rouse cooperative behavior, leading to $D_\mathrm{s} > D_\mathrm{R}$. Second, Eq.~\ref{eq:Ds_vs_Do} also predicts that $D_\mathrm{s}$ should be independent of the number of molecules, $N$, because polymer switching is a local phenomenon. The second of these predictions will be tested using the Monte Carlo method in the next Section.

\begin{figure}[!ht]
\includegraphics[width = 0.49\textwidth]{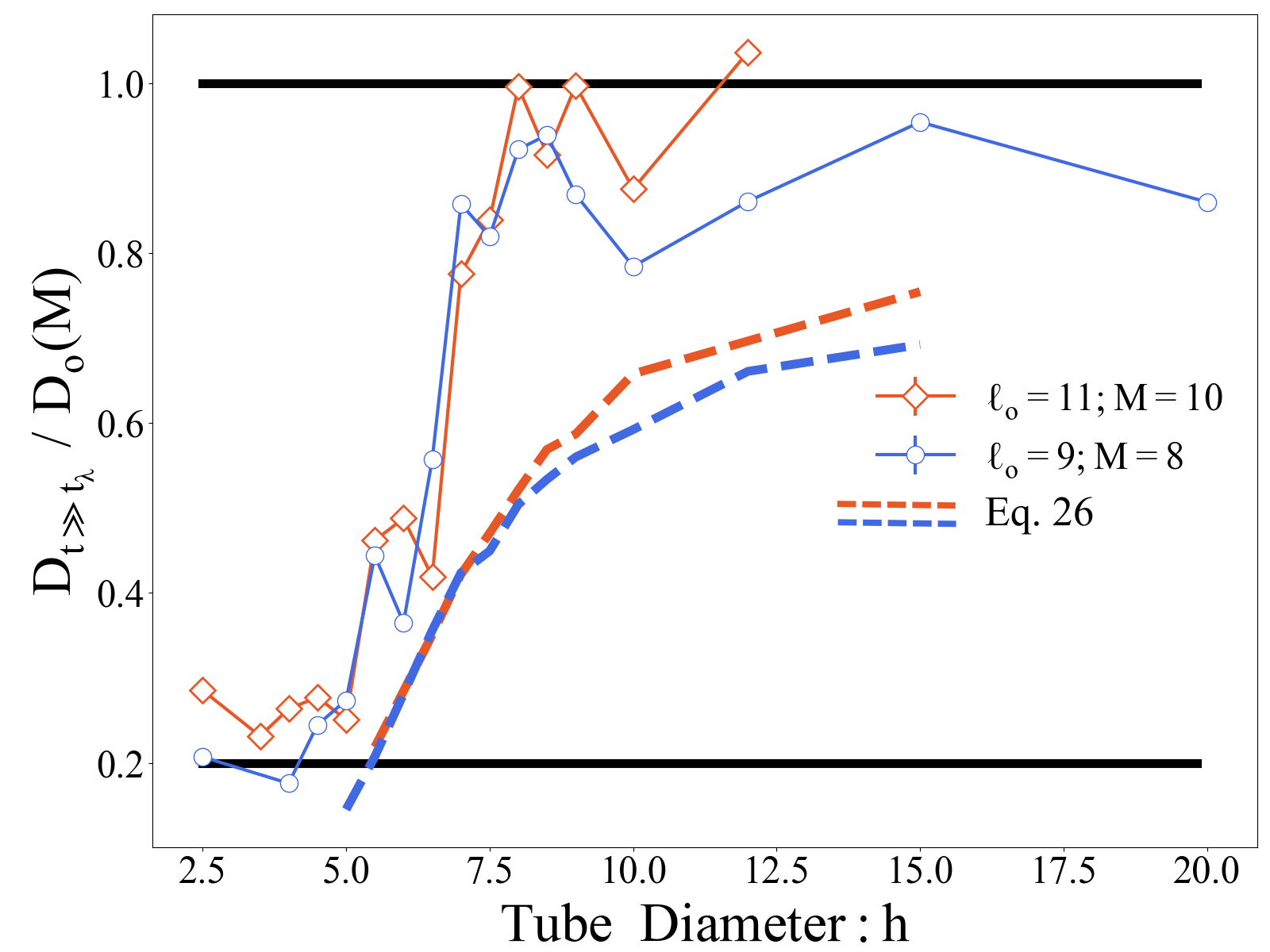}
\caption{Scaled asymptotic diffusion coefficient $D_{t\gg t_\lambda} / D_\mathrm{o}$ vs. tube diameter $h$ for $\ell_\mathrm{o}=11$, $N=5$ and $M=10$. The corresponding free diffusion coefficient is $D_\mathrm{o}(M) = \nicefrac{1}{M} = \nicefrac{1}{10}$ (shown as the upper solid line). The lower solid line gives the Rouse diffusion of the group of molecules, $D_\mathrm{R} = \nicefrac{D_\mathrm{o}}{N} = \nicefrac{D_\mathrm{o}}{5}$. The dashed line comes from Eq. \ref{eq:Ds_Do(t1)}. }
\label{fig:Diff_MD_check}
\end{figure}

The MSD lines diverge after converging at $t=t_\lambda$ in Fig.~\ref{fig:MSD_Rescale}~(b). Figure \ref{fig:Diff_MD_check} shows how the asymptotic diffusion coefficient $D_{t \gg t_\lambda}$ increases from $D_\mathrm{R}=D_\mathrm{o}/N$ to $D_\mathrm{o}$ as we increase the tube width $h$ for $N=5$ polymer chains of sizes $M=8$ and $M=10$. The diffusion coefficient increases as we transition from the Rouse regime (narrow tubes) to the free diffusion regime (wide tubes). Polymer switching occurs roughly in the narrow range $6 \le h \le 9$. We can obtain an estimate the $h$-dependence of the diffusion coefficient in this transition regime using Eq.~\ref{eq:Ds_vs_Do}, replacing $t_\mathrm{s}$ with the value of $\left<t_1\right>$:
\begin{equation}
    \frac{D_\mathrm{s}}{D_\mathrm{o}} ~\approx~ \left( \frac{8 ~ t_\mathrm{o}}{\pi \, \left<t_1\right>}  \right)^{\nicefrac{1}{2}} ~=~\frac{\ell_\mathrm{o}-S_x}{\left[ \pi \, D_\mathrm{o} \left<t_1\right>\right]^{\nicefrac{1}{2}}}  ~,
\label{eq:Ds_Do(t1)}
\end{equation}
with $S_x(h)=7.02(h-2)^{-0.47}$, $\ell_\mathrm{o}=13$ and $D_\mathrm{o}(M=10)=\nicefrac{1}{10}$. In spite of all the approximations needed for this estimate, the corresponding dashed line in Fig.~\ref{fig:Diff_MD_check} is in decent agreement with the general trend. 

\subsection{\label{subsec:MD_phase_diagram}Phase diagram}

\begin{figure}[!ht]
    \includegraphics[width = 0.49\textwidth]{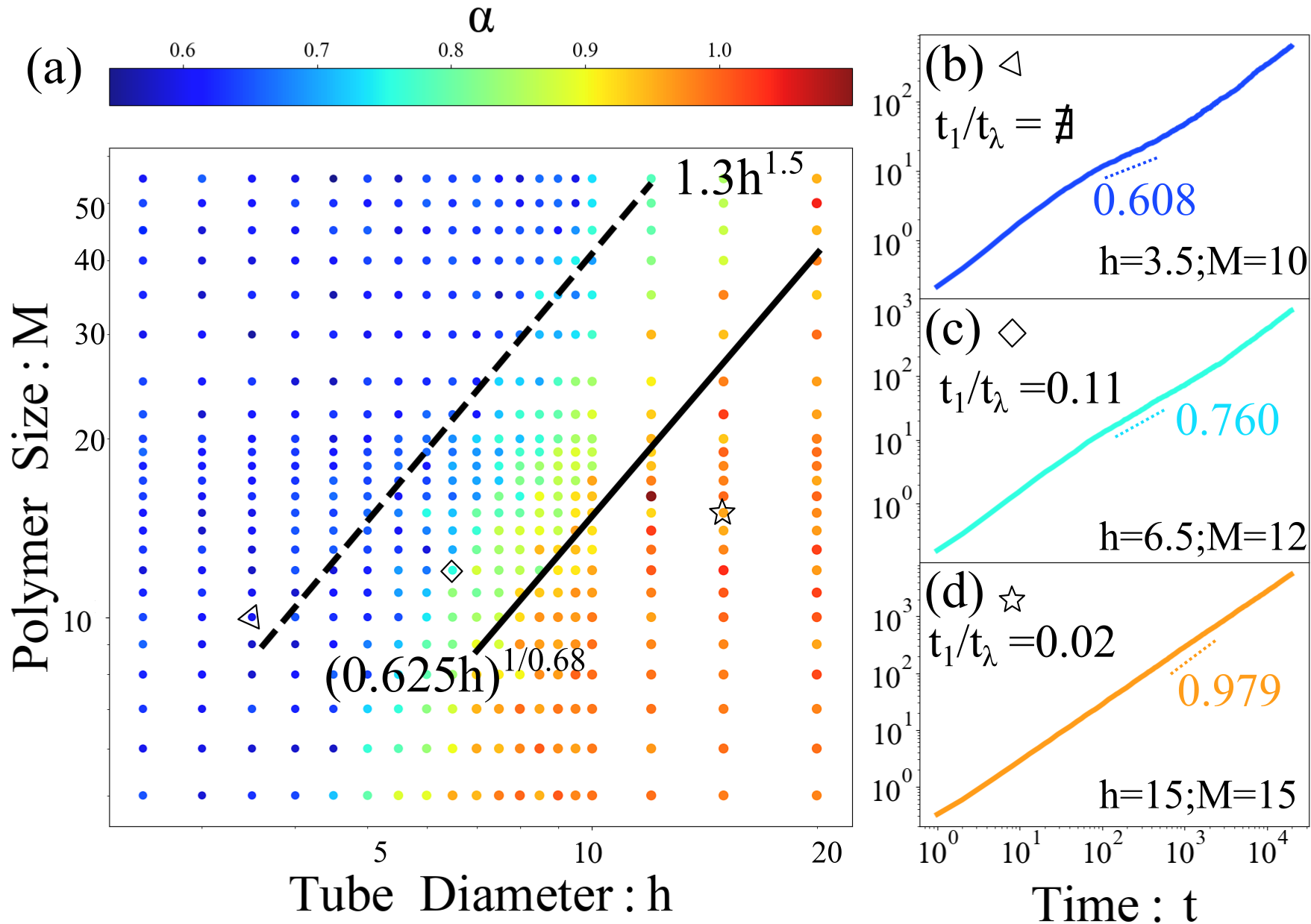}
    \caption{(a) $M-h$ phase diagram based on the value of the exponent $\alpha$ (see color code at the top) in the scaling law $\langle x^2(t) \rangle \sim t^\alpha$ when applied to intermediate times. The initial polymer separation is $\ell_\mathrm{o} = M+1$, and the number of polymers is $N=5$. The two diagonal lines are the theoretical boundaries described in the text. (b), (c) and (d) show the MSD vs time $t$ plots from which the value of the exponent $\alpha$ was obtained for the three marked points ($\triangledown$, $\diamond$ and $\star$) in (a); the ratio $t_1/t_\lambda$ is given as well.}
\label{fig:phase_figure}
\end{figure}

Figure~\ref{fig:phase_figure} (a) presents a $M-h$ phase diagram (for a system with $N=5$ chains) based on the exponent $\alpha$ found when the scaling law $\langle x^2(t) \rangle \sim t^\alpha$ is used to fit the MSD data in the SFD regime. Systems with the strongest SFD regime ($\alpha$ near $\nicefrac{1}{2}$) are located on the left, shown in blue hues. Conversely, the conditions under which SFD does not play a major role  ($\alpha$ close to 1) are found on the right (orange). 

We added two solid lines to this phase diagram. The higher one, $M=1.3\,(h-2)^{3/2}$, is suggested by Eq.~\ref{eq:blob_start}; on the left of this line the chains form strings of blobs and the switching time increases very rapidly, thereby leading to strong SFD behavior. The second line, $h = 5 \, \mathds{R}_g^\circ(M) = 1.6 \, (M-1)^{0.68}$, corresponds to a tube wide enough that two random coil chains should easily pass each other (the factor 5 is arbitrary). The points on the right of this second line correspond to cases where the polymers rarely collide with each other and essentially diffuse freely in a wide tube. Systems between these lines exhibit intermediate exponents $\alpha$ as both polymer switching and SFD coexist.

Parts (b)-(d) of Fig.~\ref{fig:phase_figure} show three examples marked by empty symbols in the phase diagram. The (b) case is in the SFD part of the phase diagram, and we observe that $\langle x^2(t) \rangle \sim t^{0.608}$ in the intermediate time regime. We were unable to observe polymer switching in this case. The (d) case is clearly in the wide tube regime, and indeed the intermediate time regime has an exponent close to unity and very short switching times. Finally, (c) is in the middle regime where both SFD and polymer switching coexist.

Next Section revisits these issues using a very efficient Monte Carlo scheme that is informed by our LD results.

\subsection{Monte Carlo Results -- Particles \textit{vs} Polymer Chains}
\label{sub:MC_results}

As described in Section~\ref{sec:LMC_methods}, our LMC simulations use periodic boundary conditions and four parameters: the number of particles, $N$, the initial spacing $\ell_\mathrm{o}$ between them, and finally the collision parameters $P$ and $Q$. As the molecular weight $M$ is irrelevant in MC simulations, all particles share the same free solution diffusion coefficient $D_\mathrm{o}$. A particle replaces a polymer chain and occupies a single lattice site. While $P$ governs the probability of chain entanglement, $Q$ dictates the disentanglement rate. For example, $P=0$ corresponds to pure SFD dynamics, the combination $[P=1,Q=0]$ leads to molecular aggregation, and we have a gas of lightly interacting chains if both parameters are large. 

As usual, we have to be careful with lattice Monte Carlo algorithms since small systems can be affected by important lattice discretization effects. On the plus side, we can now simulate larger systems, large ensembles and longer periods of time (indeed, our MC results are for ensemble sizes of 20\,000).

\begin{figure}[!ht]
\centering
\includegraphics[width = 0.48\textwidth]{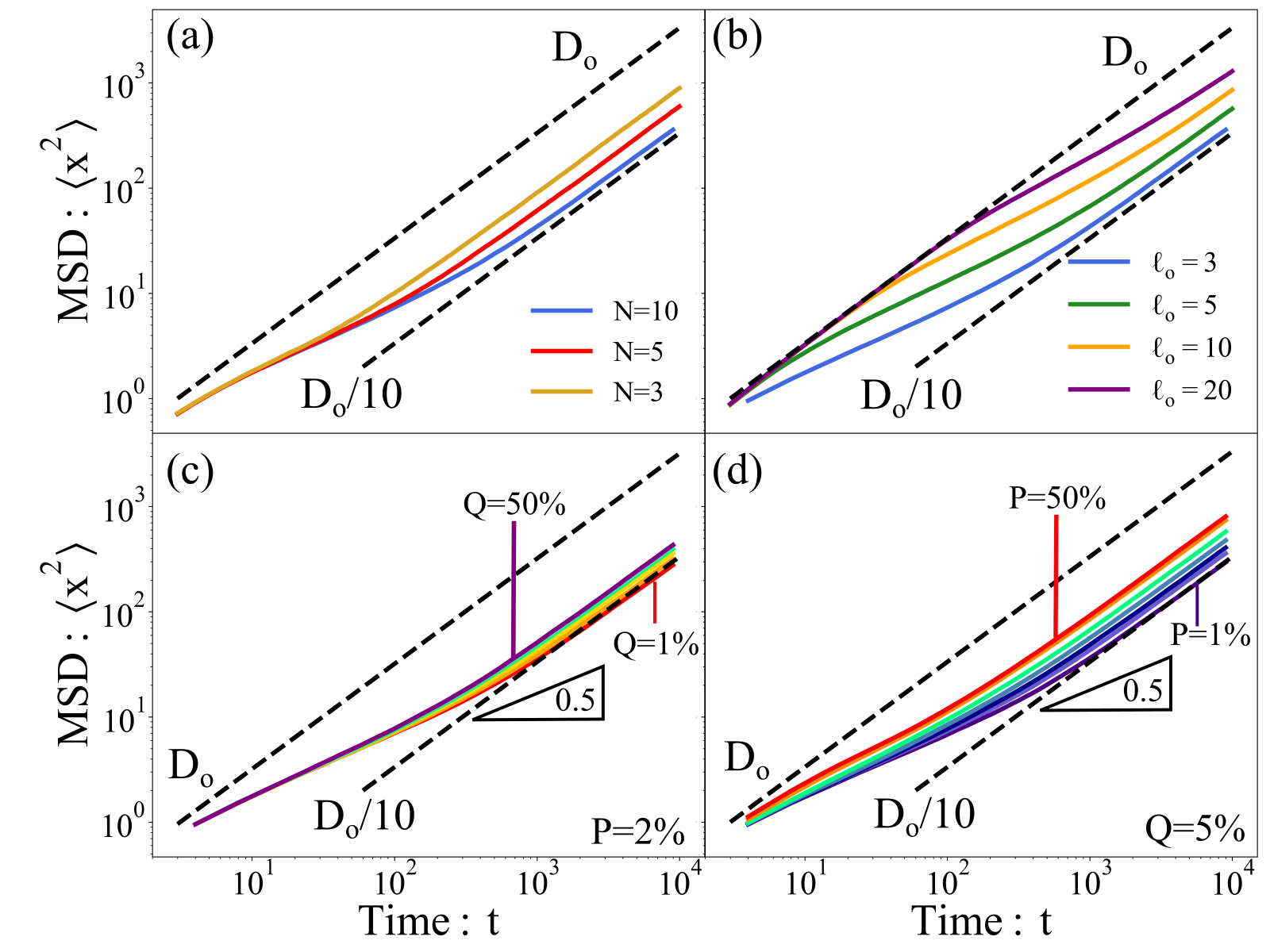}
\caption{Mean-square displacement $\left< x^2(t) \right>$ vs time $t$ for various combinations of Monte Carlo parameters. The dashed lines have a unit slope and show the theoretical limits $D=D_\mathrm{o}$ and $D=D_\mathrm{o}/10$ for the $N=10$ case. (a) $\ell_\mathrm{o} = 3$, $P=2\%$, and $Q=5\%$, for $N$ = 3, 5, and 10 particles. (b) $N = 10$, $P=2\%$, and $Q=5\%$, with mean spacings $\ell_\mathrm{o}$ = 3, 5, 10, and 20. (c) $N=10$, $P=2\%$, $\ell_\mathrm{o}=3$ and different values of $Q$. (d) $N=10$, $Q=5\%$, $\ell_\mathrm{o}=3$ and different values of $P$. }
\label{fig:MC}
\end{figure}

Figure~\ref{fig:MC} shows a range of typical simulation results (mean-square displacement $\left< x^2(t) \right>$ vs time $t$) where we systematically vary each of the four system parameters (i.e., $N$, $\ell_\mathrm{o}$, $P$, and $Q$). Panel~(a) shows the impact of the number of particles, $N$. At short times $t<t_\mathrm{o}$, the particles have yet to be affected by collisions and Eq.~\ref{eq:Short_regime} applies; indeed, the different curves then overlap and converge towards the free-solution diffusion coefficient $D_\mathrm{o}$. This is followed by a SFD regime between times $t_\mathrm{o}$ and $t_\lambda$, and a long-time regime for $t > t_\lambda$, with:
\begin{equation}
\begin{split}
 t_\mathrm{o} &= (\ell_\mathrm{o}-1)^2/8D_\mathrm{o}, \\
t_{\lambda} &= \tfrac{8}{\pi}\,N^2 t_\mathrm{o}.
\end{split}
\label{eq:MC_to_tlam}
\end{equation}
While $t_\mathrm{o} \approx 3$ is the same for all three curves, the transition to the last regime shifts to longer times $t_\lambda \sim N^2$ as $N$ increases. The $N=10$ data set almost converges with the Rouse prediction $D_\mathrm{R}=D_\mathrm{o}/10$ at long times; the small difference is due to particles being able to switch position when $P \ne 0$.

Figure \ref{fig:MC}~(b) examines the role of the initial spacing $\ell_\mathrm{o}$. The same three regimes are found, but the transition times $t_\mathrm{o}$ and $t_\lambda$ both increase like $(\ell_\mathrm{o}-1)^2$. The most interesting feature is the fact that the asymptotic diffusion coefficient slowly increases from $D_\mathrm{o}/10$ to $D_\mathrm{o}$ as we increase $\ell_\mathrm{o}$; this is due to the fact that the time between collisions scales like $t_\mathrm{o} \sim (\ell_\mathrm{o}-1)^2$ while the time required to complete a switching event is independent of the density of particles. When $\ell_\mathrm{o}$ is very large, the limiting factor is the diffusion time between collisions and we recover the free-solution diffusion coefficient $D_\mathrm{o}$. 

Figures \ref{fig:MC} (c) and (d) show how $P$ and $Q$ affect the dynamics of the particles. Large values of either $Q$ or $P$ lead to a fast switching process as long as the other parameter is not negligible. The presence of particle switching increases the asymptotic diffusion coefficient, in agreement with Eq.~\ref{eq:Ds_vs_Do}. We note that the intermediate SFD regime disappears when particles can easily change position, which is also as expected. While the different curves collapse at short times in (c), they do not in (d). This is due to the fact that this regime is controlled by the time to the first collision, $t_\mathrm{o}$: when $P$ is large, these collisions have very little or no impact on the diffusion of a particle, hence the increase towards the $D_\mathrm{o}$ limit.

\begin{figure}[!ht]
\centering
\includegraphics[width = 0.48\textwidth]{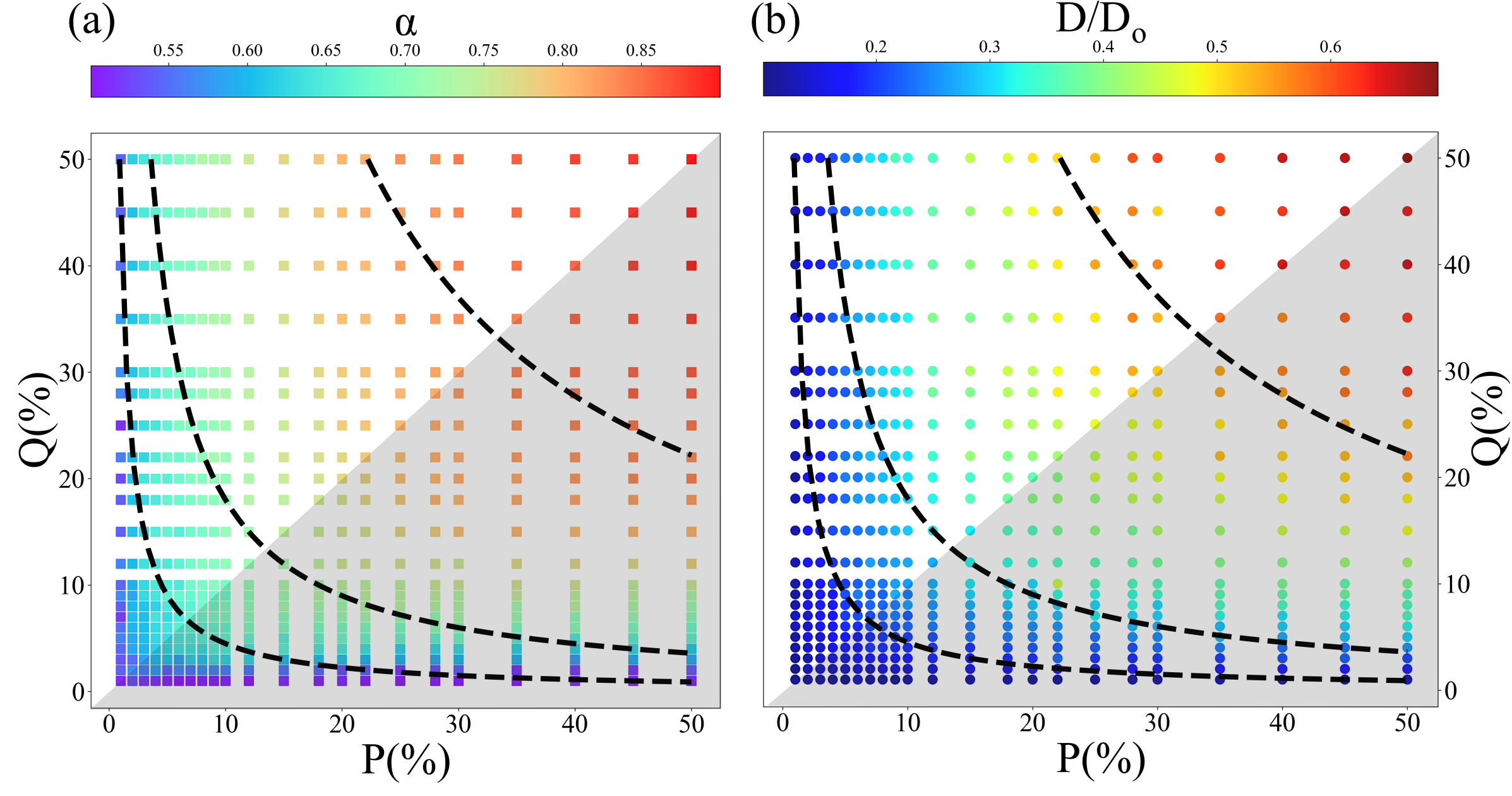}
\caption{$Q-P$ phase diagrams showing (a) the value of the SFD exponent $\alpha$, and (b) the reduced long time diffusion coefficient $D/D_\mathrm{o}$. The initial separation between the $N=10$ particles is $\ell_\mathrm{o} = 3$. The dashed lines show that the function $Q=C/P$ can be used to separate the regimes; from left to right, the values of $C$ are 45, 180, and 1110  (these choices will be explained later as they relate to Fig.~\ref{fig:Diff_MC_check_2}).}
\label{fig:MC_phase}
\end{figure}

Figure~\ref{fig:MC} shows that we can reproduce the key aspects of the physics of this polymer system with a realistic MC model. To get a better idea of the subtle impacts of $P$ and $Q$, we now present two phase diagrams. 

The first one, Fig.~\ref{fig:MC_phase} (a), describes how these parameters impact the exponent $\alpha$ in the SFD regime and is thus the MC equivalent to Fig.~\ref{fig:phase_figure} (a). Strikingly, there is a nearly symmetric distribution of points, with the value of the exponent $\alpha$ growing with both $P$ and $Q$, producing a rainbow effect. The reason for this comes from the MC rules described in Sec. \ref{sec:LMC_methods}; if we consider only two-particle interactions and neglect the diffusion of particle aggregates, the duration of an event during which two particles change position is $t_\mathrm{s} \sim 1/PQ$. The $Q \sim 1/P$ dashed lines separate the three regimes, just like the two straight lines did in Fig.~\ref{fig:phase_figure}~(a). The second phase diagram, Fig.~\ref{fig:MC_phase} (b), shows the impact of these parameters on the particles' asymptotic diffusion coefficient. We again observe the quasi-symmetry along the diagonal. 

Together, these two phase diagrams demonstrate that SFD survives if particle-particle collisions rarely lead to switching events. SFD "survival" means that we keep these two elements: 1) the exponent $\alpha$, which describes the MSD in the SFD regime $\left< x^2(t) \right> \sim t^\alpha$, is close to $\nicefrac{1}{2}$; 2) the asymptotic diffusion coefficient of the particles is close to $D_\mathrm{o}/N$ because SFD leads to cooperative motion of the particles at long times. Both phase diagrams clearly show these two limits.

Note that we shadowed the $P>Q$ areas in both phase diagrams as we consider this situation to be  unrealistic. In this hypothetical region, particles (polymer chains) would have a high probability to overlap ($P$) and a low probability to separate ($Q$), which seems to make little sense unless there are attractive interactions leading to stable aggregates, a situation that is beyond the scope of this paper.

\begin{figure}[!ht]
\centering
\includegraphics[width = 0.48\textwidth]{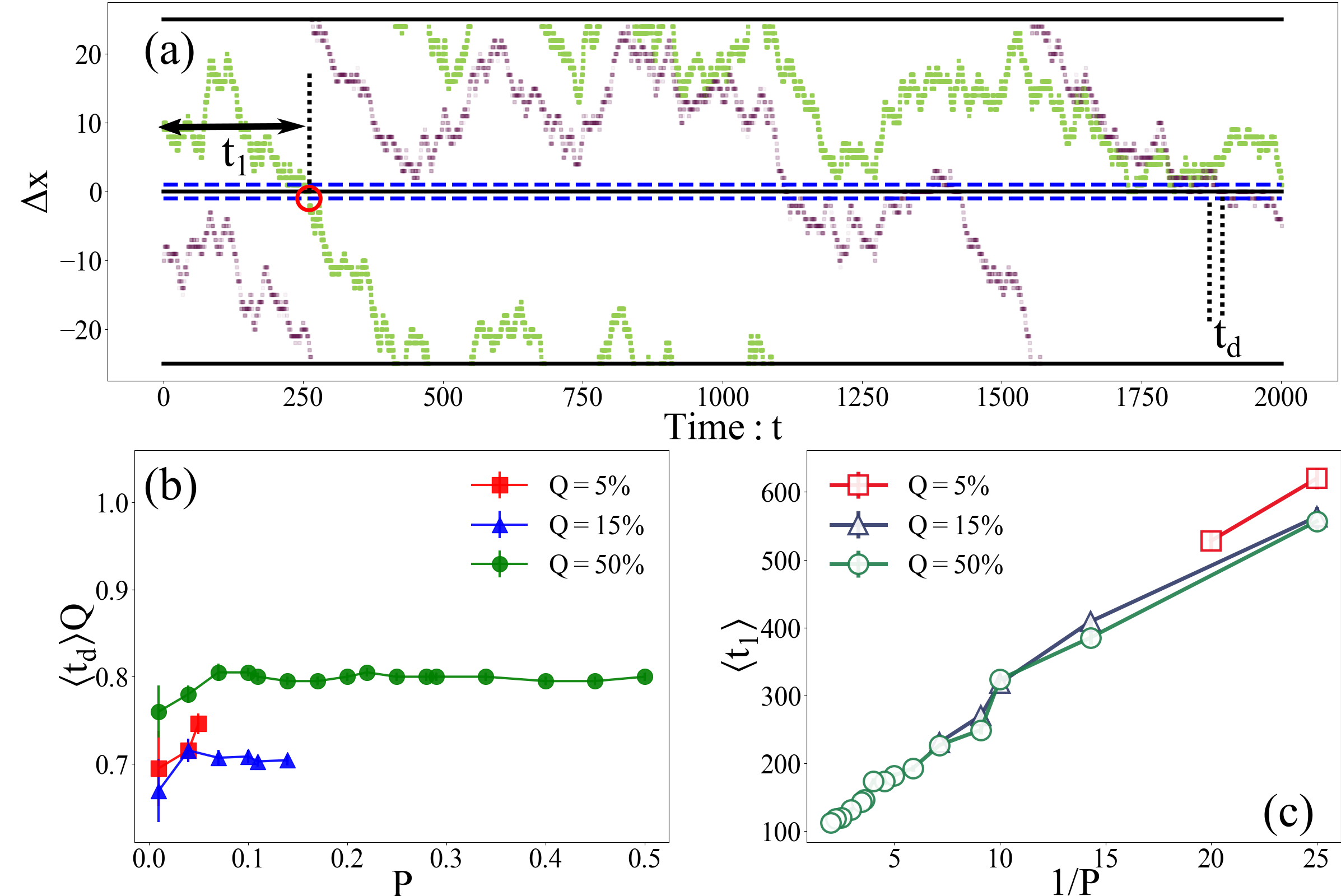}
\caption{(a) Simulation examples showing the time dependence of the distance $\Delta x(t)$ between two pairs of nearest neighbor particles. The system parameters are $P=12\%$, $Q=15\%$, $\Delta x(0)=\ell_\mathrm{o}=10$, $N=5$. The first particles to change positions do so at time $t_1=260$. The duration of the event starting at time $t=1885$ is $t_\mathrm{d} = 19$. (b) Scaled mean event duration $\langle t_\mathrm{d} \rangle Q$ vs $P$. (c) Mean first switching time $\langle t_1 \rangle$ vs $1/P$. The ensemble size for B and C is 1\,000.  }
\label{fig:MC_switch}
\end{figure}

In order to investigate the switching mechanism, we repeated the study that led to Fig.~\ref{fig:switching_figure} using our MC algorithm. An example is shown in Fig.~\ref{fig:MC_switch} (a). From these trajectories, we extracted the time to the first switching event, $\left< t_1 \right>$ and the mean duration of successful switching events, $\left< t_\mathrm{d} \right>$ (both are defined in panel A). 
Note that we will focus on the $Q \geq P$ part of the phase diagram, for the reasons given above. 

Figure~\ref{fig:MC_switch} (b) shows that the mean duration of successful switching events, $t_\mathrm{d}$, is essentially independent of the parameter $P$ but inversely proportional to $Q$. This is as expected since $t_\mathrm{d}$ is in fact the disentanglement time of two particles, a process that is controlled by $Q$ in our MC algorithm; the small residual $Q$ and $P$ dependencies come from multi-particle interactions. Figure~\ref{fig:MC_switch} (c) shows that the mean first switching time $\langle t_1 \rangle$ grows linearly with $1/P$, while exhibiting a very weak dependence on $Q$. This result is also expected since $P$ controls the first switching event in the MC algorithm. As $t_\mathrm{d} \sim \nicefrac{1}{Q}$ and $t_1 \sim \nicefrac{1}{P}$, we can conclude that $t_\mathrm{s}$, which includes both of these processes, should scale roughly like $t_\mathrm{s} \sim \nicefrac{1}{PQ}$, in agreement with the structure of the phase diagrams in Fig.~\ref{fig:MC_phase}. 

\begin{figure}[!ht]
\centering
\includegraphics[width = 0.48\textwidth]{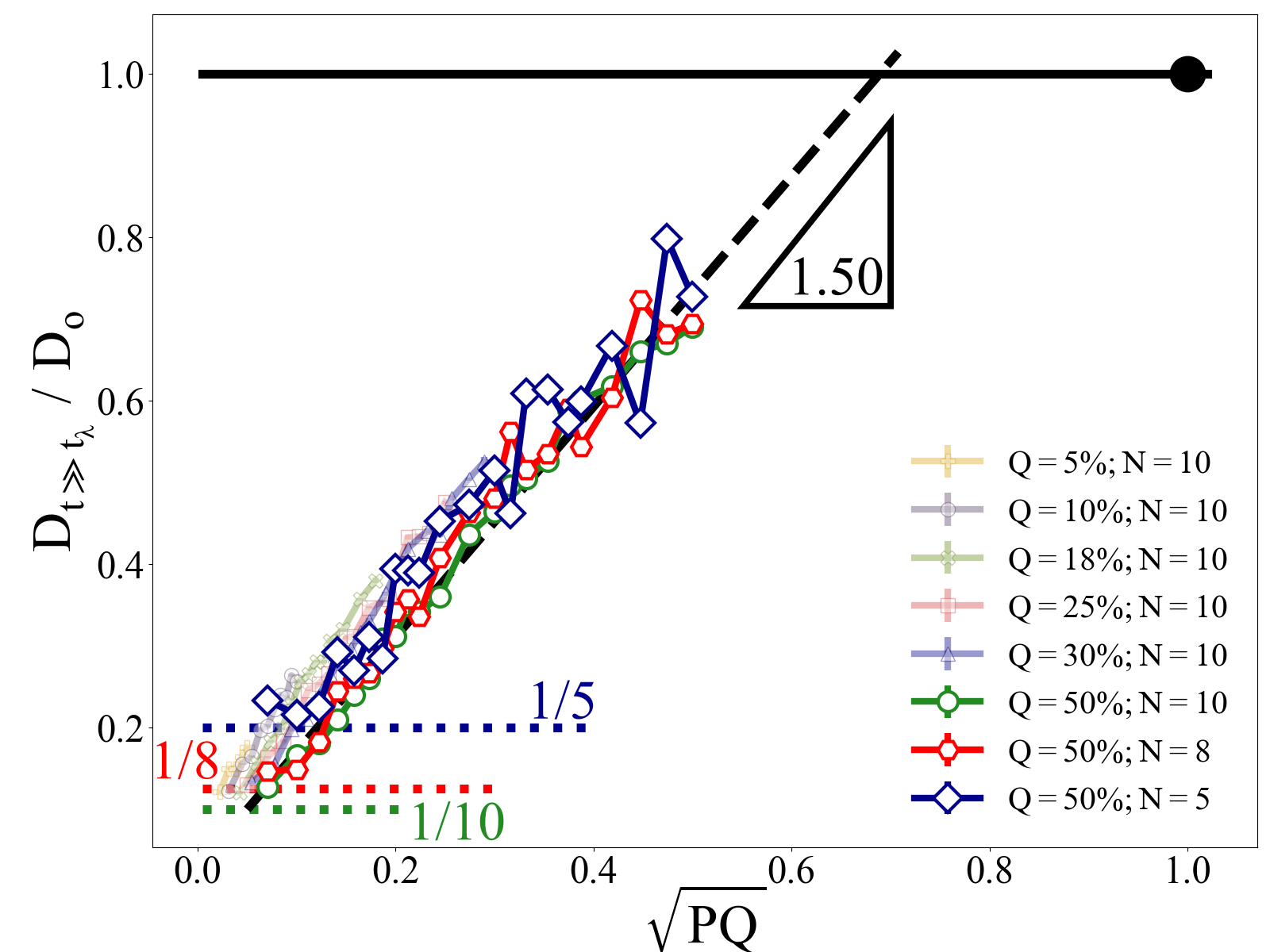}
\caption{Scaled asymptotic diffusion coefficient $D_{t\gg t_\lambda} / D_\mathrm{o}$ vs. $\sqrt{PQ}$ for various combinations of $Q$ and $N$; the system parameters are on the figure, and the fixed separation  is $\ell_o=3$. The lower dotted lines give the Rouse prediction $D_\mathrm{R} = \nicefrac{D_\mathrm{o}}{N}$ for the group, while the solid line at the top (in black) shows the free-solution diffusion coefficient $D_\mathrm{o}$; the (1,1) point is the theoretical free-diffusion case. The dashed line is a linear fit for the $N=10$ and $Q=50\%$ data set.}
\label{fig:Diff_MC_check_2}
\end{figure}

Equations~\ref{eq:Ds_vs_Do} and \ref{eq:Ds_vs_DR} indicate that the diffusion coefficient scales as $D \sim t^{-\nicefrac{1}{2}}_\mathrm{s}$ when the switching time falls within the range $t_\mathrm{o} < t_\mathrm{s} < t_\lambda$. This transition regime is shown in Fig.~\ref{fig:Diff_MD_check}, where $D$ rises from the Rouse value $D_\mathrm{R}=D_\mathrm{o}/N$ (when $t_\mathrm{s} > t_\lambda$) to the free diffusion value $D=D_\mathrm{o}$ (when $t_\mathrm{s} <t_\mathrm{o}$) as polymer switching frequency increased ($h$ was increased). Since we have roughly $t_\mathrm{s} \sim 1/PQ$ in our MC algorithm, we should expect $D \sim \sqrt{PQ}$ here. Figure~\ref{fig:Diff_MC_check_2} confirms that this is the case in the intermediate regime for various parameter combinations. The data points approach the $D_\mathrm{R}=D_\mathrm{o}/N$ limit as $PQ$ decreases. Because we restricted our parameters to $P \le Q \le \nicefrac{1}{2}$, the product $PQ$ is not large enough to reach the free-diffusion limit $D=D_\mathrm{o}$. The theoretical $D(P\!=\!Q\!=\!1) = D_\mathrm{o}$ point on the top line demonstrates that the linear intermediate regime does not extend to this point. Our MC model is unsuitable for modeling the regime where particles barely interact (corresponding to $h \gg \mathds{R}_g^\circ(M)$ for polymer chains).

The dashed line in Fig.~\ref{fig:Diff_MC_check_2}  is the linear fit $D/D_\mathrm{o} = 1.50(2) \sqrt{PQ}$ to the $[N\,=\,10; Q\,=\,50\%]$ data set. The fact that this fit also agrees with the other data sets demonstrates that the diffusion coefficient does not depend on the number of particles $N$ in this intermediate regime, an important prediction of  Eq.~\ref{eq:Ds_vs_Do}.  For the $N=10$ case, this fit predicts that $D=D_\mathrm{R}=D_\mathrm{o}/10$ when $Q \approx 0.45/P$, which should be interpreted as marking the hard limit for SFD dynamics (no particles exchanging position); similarly, it predicts that $D=D_\mathrm{o}/2$ when $Q \approx 11.1/P$, which should be the point beyond which particle switching dominates. Finally, it gives $D=D_\mathrm{R}=2D_\mathrm{o}/10$ when $Q \approx 1.8/P$, possibly pointing to a condition where SFD and particle switching play equal roles. These three lines are shown in the phase diagrams, Fig.~\ref{fig:MC_phase}; we can clearly see that they indeed correspond to the description given above.



\section{Discussion and Conclusion}
\label{sec:Conclusion}

We investigated the dynamics of one-dimensional dilute polymer solutions in a closed continuous system resembling a toroidal tube, a configuration that is amenable to experimental study in the laboratory. The polymer molecular weight $M$ and tube width $h$ influence the polymers' ability to swap positions. The main theoretical predictions regarding the different regimes to be expected were presented, and two different simulation methods were used to model the system: Langevin dynamics and Monte Carlo, with the aim to establish the link between these approaches. 

The Monte Carlo algorithm was designed to minimize the number of parameters (we used only two, denoted $P$ and $Q$) while properly accounting for the various features of the polymer systems, such as the entanglement and disentanglement processes. The $Q$ parameter is actually unique to polymer systems since it is decoupled from $P$. For instance, it could allow us to study aggregating systems. In practice, we restricted the study to $Q \ge P$, which seems to be realistic for non-interacting polymer chains in good solvents.

The LD study confirmed the existence of the many time regimes predicted in the theory section, namely free diffusion, single file diffusion, polymer switching, and Rouse regimes. Both the tube width $h$ and the polymer size $M$ play a role in determining whether polymer switching will displace the Rouse regime at long times. In fact, the key parameter is the number of blobs formed by the polymer chains in the tube, since chains with multiple blobs are unlikely to interpenetrate and change position on a time scale that would alter the long-term behavior of the polymer solution. Consequently, the tube diameter range where polymer switching plays a role is relatively narrow. 

Polymer chains changing position compete with single file dynamics at intermediate times and with Rouse dynamics at longer times. We therefore proposed to use the $\alpha$ exponent, which describes the time evolution of the mean square displacements (MSD) at intermediate times, as a way of characterizing the effect of polymer switching. This has led to the $M-h$ phase diagram in Fig.~\ref{fig:phase_figure}, where we can see that the system transitions smoothly between the different regimes predicted by the theory. However, modeling these systems with LD is challenging due to the long simulation times needed to gather sufficient polymer switching events for statistical analysis. For this reason, we have also had a look at this problem with the help of an MC algorithm.

The results of the MC method are in agreement with the LD data and theoretical predictions. We thus proposed two phase diagrams in Fig.~\ref{fig:MC_phase} based on the $\alpha$ exponent and the asymptotic diffusion coefficient. Importantly, Fig.~\ref{fig:Diff_MC_check_2} validated the theoretical description of the competition between SFD and particle exchange. We thus conclude that the MC algorithm can be used to study this class of problems.

\begin{figure}[!ht]
\centering
\includegraphics[width = 0.48\textwidth]{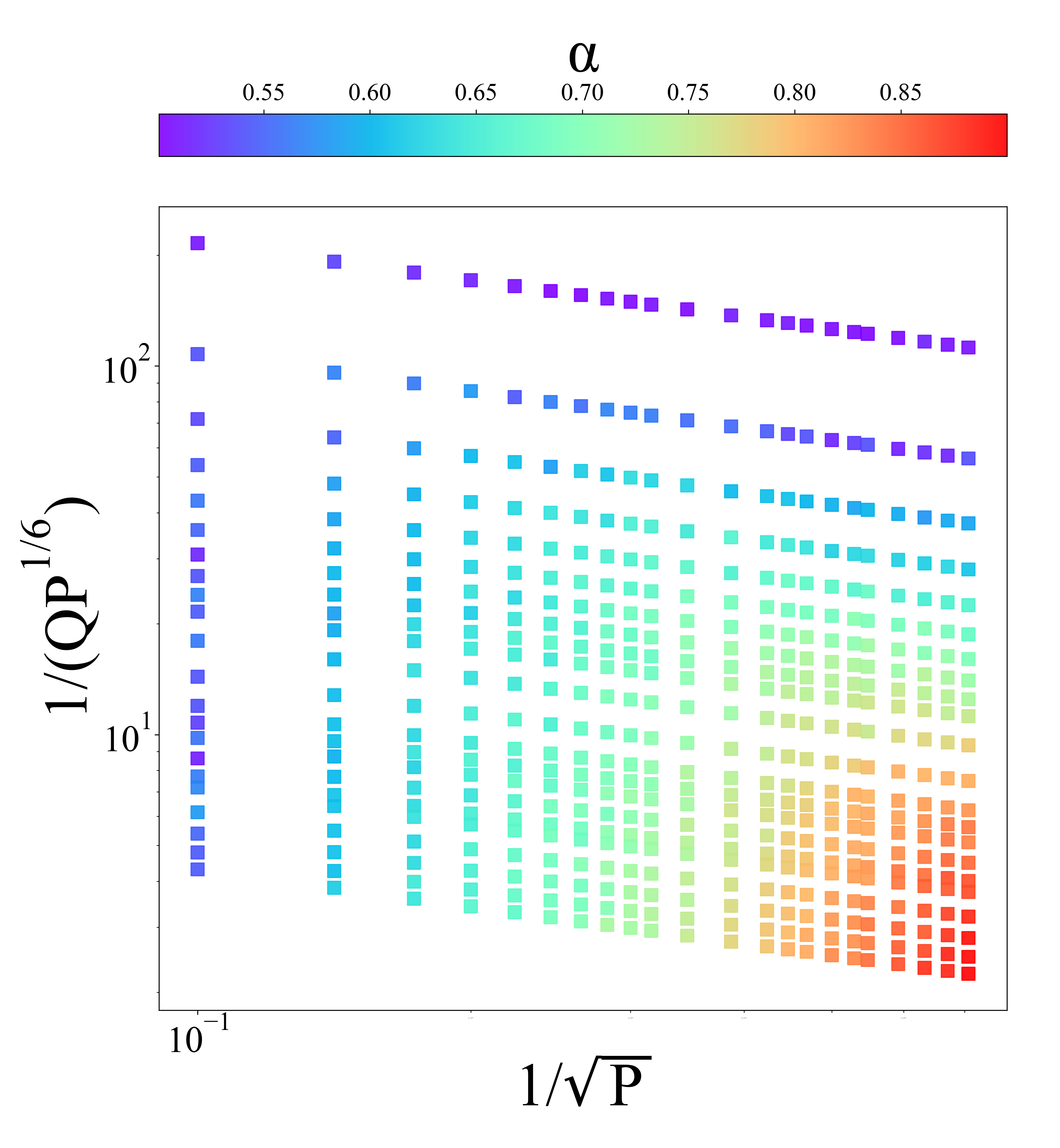}
\caption{Phase diagram presented in Fig.~\ref{fig:MC_phase}~(a) with both axes replaced by the variables defined in Eq.~\ref{eq:hMPQ}. }
\label{fig:Redo_MC_phase}
\end{figure}

In order to connect the LD and MC parameters, $[M,h]$ and $[P,Q]$, respectively, we first use the fact that the disentanglement time of two overlapping chains in a tube\cite{ref:Arnold} scales like $t_d \sim M^2 h^{2-1/\nu}$. The MC result $t_d \sim 1/Q$ implies that $Q \sim 1/M h^{2-1/\nu}$, where we removed one power of $M$ because the duration of one MCS scales as $\tau_c\sim M$. We then note that while the different regimes in the LD phase diagram are separated on the basis of the number of blobs (hence the dashed diagonal lines $Mh^{-1/\nu}=constant$), these regimes are separated on the basis of the combined parameter $1/PQ=constant$ in the MC phase diagrams. We will thus assume that $Mh^{-1/\nu} \sim 1/PQ$. Using these two relations, we get
\begin{equation}
\begin{split}  
   h &\sim \sqrt{P} \\   M &\sim 1/QP^{1-1/2\nu} \sim 1/QP^{\nicefrac{\,1}{6}}~,
\end{split} 
\label{eq:hMPQ}
\end{equation}
where the last result is obtained using Flory's exponent $\nu=3/5$. Figure \ref{fig:Redo_MC_phase} shows the phase diagram presented in Fig.~\ref{fig:MC_phase}~(a) with the two axes modified using these two expressions. The similarity with Fig.~\ref{fig:phase_figure} (a) suggests that it is indeed possible to connect the two simulation methods. The approach used here is not the only option; further research will be necessary in order to fully exploit the MC methods and obtain a better connection between the parameters of the two models.

The polymer SFD problem studied in this paper suggests a number of interesting extensions. For example, it would be interesting to study how the regimes described in our work would evolve as the concentration of the polymers is increased towards the semi-dilute limit. On a different note, it would also be of interest to study bimodal polymer solutions (e.g., linear and ring polymers with identical molecular weights $M$) to see if the different switching times can lead to SFD-triggered phase separation along the torus. 

As for our MC approach, it can be modified to study a much broader range of one-dimensional problems, including bimodal (or polydisperse) dilute polymer solutions and polymer aggregation. Adding a bias to the movement of one of the lattice particles could also allow us to investigate one-dimensional electrophoresis \cite{ref:Collision_paper}.

Our investigations used LD and MC simulations, as well as simple analytical theory. These analytical and computational methods did not account for the influence of hydrodynamic interactions and polymer rigidity, in addition to various other factors. While hydrodynamics likely wouldn’t qualitatively alter the results, polymer stiffness might, as the Kuhn length introduces a new scale, potentially leading to new regimes.

\section{\label{sec:Acknowledgement}Acknowledgements}
{The LD simulations were performed with the ESPResSo package, and we made use of the computing resources provided by SHARCNET and Compute Canada. GWS acknowledges the support of both the University of Ottawa and the Natural Sciences and Engineering Research Council of Canada (NSERC), funding reference number RGPIN/046434-2013 (Discovery Grant).}

\nocite{*}
\bibliography{sfd_draft}
\end{document}